\documentclass{PoS}

\def\asec{\ifmmode ^{\prime\prime}\else$^{\prime\prime}$\fi}

\def\degs{\ifmmode ^{\circ}\else$^{\circ}$\fi}
\def\amin{\ifmmode ^{\prime}\else$^{\prime}$\fi}
\def\asec{\ifmmode ^{\prime\prime}\else$^{\prime\prime}$\fi}

\def\degs{\ifmmode ^{\circ}\else$^{\circ}$\fi}
\def\amin{\ifmmode ^{\prime}\else$^{\prime}$\fi}

\def\cm{\mbox{\,cm}}

\def\cm3{\mbox{\,cm$^{-3}$}}

\def\lsim{\!\!\!\phantom{\le}\smash{\buildrel{}\over
 {\lower2.5dd\hbox{$\buildrel{\lower2dd\hbox{$\displaystyle<$}}\over
                                 \sim$}}}\,\,}
\def\gsim{\!\!\!\phantom{\ge}\smash{\buildrel{}\over
{\lower2.5dd\hbox{$\buildrel{\lower2dd\hbox{$\displaystyle>$}}\over
                               \sim$}}}\,\,}

\title{High-resolution radio observations of nuclear and circumnuclear starbursts in Luminous Infrared Galaxies}

\ShortTitle{High-resolution radio observations of (circum)nuclear starbursts}

\author{\speaker{Miguel \'Angel P\'erez-Torres} and Antonio Alberdi \\
        Instituto de Astrofísica de Andalucía - CSIC, E-18080 Granada, Spain\\
        E-mail: \email{torres@iaa.es}, 
                \email{antxon@iaa.es}}

\abstract{ 
High-resolution radio observations of nearby starburst
galaxies have shown that the distribution of their radio emission consists
of a compact ($\leq$ 150 pc), high surface brightness (T$_b\geq$
10$^3$ K) central radio source immersed in a low surface brightness
circumnuclear halo. This radio structure is similar to that detected
in bright Seyferts galaxies like NGC~7469 or Mrk~331, which display
clear circumnuclear rings.  While the compact, centrally
located radio emission in these starbursts might be generated by a
point-like source (AGN), or by the combined effect of multiple radio
supernovae and supernova remnants (e.g., the evolved nuclear starburst
in Arp~220), it seems well established that the circumnuclear regions
of those objects host an ongoing burst of star-formation (e.g.,
NGC~7469; \cite{col01}, \cite{alb06}).
Therefore, high-resolution radio observations of Luminous Infra-Red Galaxies
(LIRGs) in our local universe are a powerful tool to probe the
dominant dust heating mechanism in their nuclear and circumnuclear
regions.  

In this contribution, we show results obtained from VLA-A, MERLIN, and
EVN (VLBI) radio observations of the galaxies NGC 7469 ($D\approx 70$
Mpc) and IRAS 18293-3413 ($D\approx 79$ Mpc), where two extremely
bright radio supernovae have been found.  High-resolution studies of
these and other LIRGs would allow us to determine the core-collapse
supernova rate in them, as well as their star-formation rate.  }

\FullConference{From planets to dark energy: the modern radio universe\\
		 October 1-5 2007\\
		 University of Manchester, Manchester, UK}

\begin{document}

\section{Introduction}

Galaxies at the highest infrared luminosities ($L{_{\rm
IR}}[8-1000~{\mu}m] \geq 10^{12} L{_\odot}$), known as Ultra-Luminous
Infrared Galaxies (ULIRGs), appear to be advanced merger systems and
may represent an important stage in the formation of quasi-stellar
objects.  The bulk of the energy radiated by ULIRGs is infrared
emission from warm dust grains heated by a central power source.
The critical question concerning these galaxies is whether
the dust in the central regions ($r \lsim$1~kpc) is heated by a
starburst or an active galactic nucleus (AGN), or a combination of
both. Mid-infrared spectroscopic studies of ULIRGs suggest that the
vast majority of these galaxies are powered predominantly by recently
formed massive stars, with a significant heating from the AGN only in
the most luminous objects \cite{vei99}. These authors 
also found that at least half of ULIRGs are probably powered by
both an AGN and a starburst in a circumnuclear disk or ring.  These
circumnuclear regions are located at radii $r\simeq$700~pc from the
nucleus of the galaxy, and also contain large quantities of dust.

High-sensitivity, high-resolution radio observations can prove to be
extremely important in studying both the nuclear and circumnuclear
regions of ULIRGs, as radio emission is not affected by dust
extinction, and the use of VLBI techniques allows for parsec, or even
sub-parsec resolution.  Indeed, \cite{col01} have discovered a
bright radio supernova (SN~2000ft) in the circumnuclear starburst region of the
Highly Luminous IR Galaxy NGC~7469, and the most spectacular evidence
of a compact starburst in a ULIRG is the discovery of a population of
bright radio supernovae and supernova remnants in the nuclear regions
of Arp~220 \cite{smi98} and Mrk~273 \cite{bon05}, using VLBI observations.

\section{VLA, MERLIN, and VLBI observations of (circum)nuclear starbursts}

In Figure 1, we show a few examples where high-resolution radio
observations have been extremely useful in unveiling dust-enshrouded
supernovae in circumnuclear starbursts, as SN~2000ft in NGC~7469
(panel a; see also \cite{col01} and \cite{alb06}), or SN~2004ip in
IRAS 18293-3413 (panel c and \cite{per07}).  SN 2000ft was first
discovered at radio wavelengths \cite{col01}, and therefore its case
nicely shows that high-resolution radio observations of
(circumn)nuclear starbursts are crucial to unveil supernovae in
dust-enshrouded environments.  Further, the monitoring of SN 2000ft
showed that it is likely a type IIn event \cite{alb06}, like e.g., SN
1986J in NGC 891 \cite{per02}, i.e., and behaves in much the way
standard radio supernovae evolve.  We also show in panel (b) 18 cm
high angular resolution radio observations of the innermost $\sim
200$pc of the nuclear region of NGC~7469 (panel b) imaged with MERLIN
(top) and the EVN (bottom).  The EVN image shows a number of compact
regions, whose exact nature is still under discussion, and will be
presented elsewhere.

It is foreseeable that the advent of the SKA will permit to carry out
such studies for a large sample of LIRGs in the local universe. This
should allow us to settle the question of the dominant heating
mechanism in LIRGs and ULIRGs in the local universe. In addition, the
enormous sensitivity of the array would allow to fill in the existing
gap between the young supernova phase and its late, supernova remnant
phase, by permitting to image historical supernovae that cannot be
currently imaged due to limiting sensitivity of the existing VLBI
arrays.

\begin{figure}
\includegraphics[width=\textwidth]{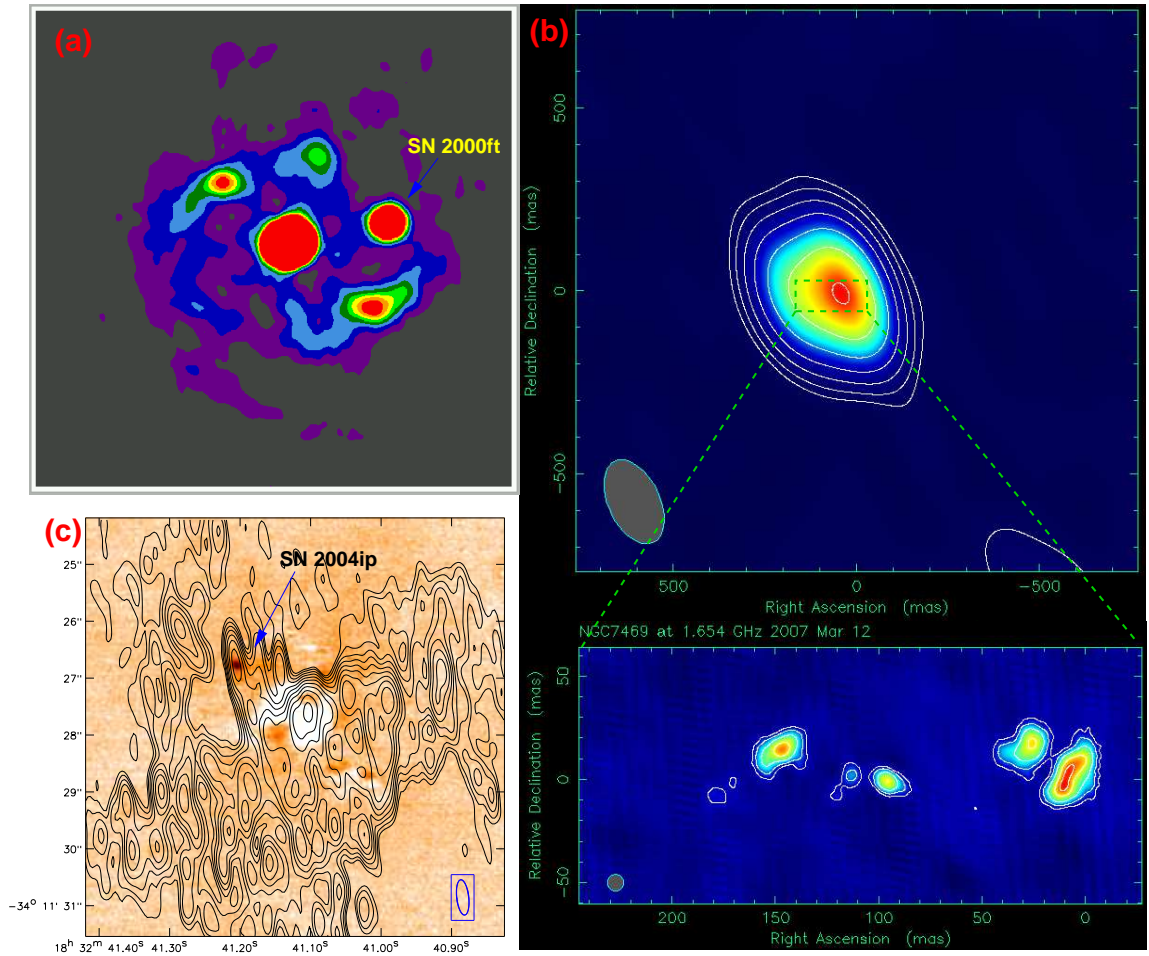}
\caption{Panel (a): 8.4 GHz VLA discovery image of the radio supernova
SN~2000ft in the galaxy NGC~7469 \cite{col01}; panel (b): 1.7 GHz
simultaneous, MERLIN (top) and EVN (bottom) images of the nucleus of
galaxy NGC~7469; note that the EVN image
clearly discerns the substructure tha is not resolved by MERLIN
(Alberdi et al. in preparation); panel (c): 8.4 GHz VLA image of SN
2004ip, whose detection at radio wavelengths three years after its
explosion confirmed it core-collapse nature \cite{per07}. }
\label{figure}
\end{figure}


\begin{thebibliography}{99}
\bibitem{alb06} A.~Alberdi et al.,  2006, ApJ, 638, 938
\bibitem{bon05}M.~Bondi, M.A.~P\'erez-Torres, D. Dallacasa \& T.W.B Muxlow, 2005, {\emph MNRAS}, 361, 748
\bibitem{col01}L.~Colina et al., 2001, {\emph ApJ}, 553, L19
\bibitem{per02}M.A.~P\'erez-Torres et al., 2002, {\emph MNRAS}, 335, L23
\bibitem{per07}M.A.~P\'erez-Torres et al., 2007, {\emph ApJ}, 671, L21
\bibitem{smi98}H.E. Smith  et al. 1998, ApJ, 493, L17
\bibitem{vei99}S. Veilleux, Sanders \& Kim, 1999, {\emph ApJ}, 522, 113

\end{thebibliography}
\end{document}